\documentclass[
 aip,       % one, two column
 amsmath,amssymb,
 twocolumn,  % two column
 reprint,   % one column
]{revtex4}

\usepackage{textcase}
\usepackage{graphicx}% Include figure files
\usepackage{dcolumn}% Align table columns on decimal point
\usepackage{bm}% bold math
\usepackage{amsmath,amsthm,amssymb,mathrsfs}

\begin{document}

\title{Current Dependence of Spin Torque Switching Barrier}

\author{Tomohiro Taniguchi}
%\email{tomohiro-taniguchi@aist.go.jp}
\author{Hiroshi Imamura}
%\email{h-imamura@aist.go.jp}

\affiliation{ 
National Institute of Advanced Industrial Science and Technology (AIST), Spintronics Research Center, 1-1-1 Umezono, Tsukuba 305-8568, Japan
}

\date{\today}% 

\begin{abstract}
  The current dependence of the switching barrier 
  for spin torque switching of an in-plane magnetized ferromagnet 
  was studied. 
  Two scaling currents, $I_{\rm c}$ and $I_{\rm c}^{*}(>I_{\rm c})$, were introduced 
  to distinguish the magnetization stability. 
  In the low-current region $I<I_{\rm c}$, 
  the switching barrier is linear to the current 
  with another scaling current $\tilde{I}_{\rm c}$, 
  while such linear scaling does not hold in the high-current region $I_{\rm c} \le I < I_{\rm c}^{*}$. 
  The linear scaling is valid for the high temperature and the long current pulse duration time. 
\end{abstract}

\maketitle

%======================================================================
% Introduction
%======================================================================

Spin torque switching of a nanostructured ferromagnet has been extensively studied 
because the switching probability in a thermally activated region 
provides us important information about 
spintronics devices such as 
the thermal stability of 
magnetic random access memory (MRAM) \cite{albert02,myers02,lacour04,yagami05,morota08,yakata09,bedau10,heindl11,sukegawa12}. 
Since the retention time of MRAM depends strongly on its thermal stability, 
an accurate evaluation of the thermal stability is required. 
However, there is some controversy regarding 
the theoretical formula of the switching probability 
of an in-plane magnetized system \cite{koch04,li04,apalkov05,wang08,swiebodzinski10,pinna13,taniguchi12b,taniguchi13,taniguchi13a}. 
The issue is the validity of the linear scaling 
of the current dependence of the switching barrier. 
Here, the switching barrier $\Delta$ relates to the switching probability $P$ 
through the switching rate $\nu=fe^{-\Delta}$ as $P=1-e^{-\nu t}$, 
in which $f$ and $t$ are the attempt frequency and current pulse duration time, respectively. 
The thermal stability can be defined as 
the switching barrier in the absence of a current. 
In the analyses of the experiments, 
the switching barrier is assumed to be 
\begin{equation}
  \Delta
  =
  \Delta_{0}
  \left(
    1
    -
    \frac{I}{I_{\rm c}^{*}}
  \right)^{b},
  \label{eq:Delta_general}
\end{equation}
where the thermal stability $\Delta_{0}=MH_{\rm K}V/(2k_{\rm B}T)$ depends on 
the magnetization $M$, 
the uniaxially anisotropic field along the in-plane easy axis $H_{\rm K}$, 
the volume of the free layer $V$, 
and the temperature $T$. 
The current is denoted as $I$ 
while $I_{\rm c}^{*}$ is 
the spin torque switching current at zero temperature, 
by which the thermally activated region is defined as $I < I_{\rm c}^{*}$. 
The important point is that 
the switching exponent $b$ 
in previous works \cite{koch04,li04,apalkov05,wang08,swiebodzinski10} was assumed to be unity. 
Since those publications, 
linear scaling ($b=1$) has been widely used to analyze 
the spin torque switching experiments \cite{lacour04,yagami05,morota08,yakata09,bedau10,heindl11,sukegawa12}. 
On the other hand, 
refs. \cite{taniguchi12b,taniguchi13} argued that 
the switching exponent $b$ depends on the current. 
The value of $b$ is larger than unity 
in the relatively low-current region $I \ll I_{\rm c}^{*}$, 
and reaches almost square in the relatively high-current region, $I \lesssim I_{\rm c}^{*}$. 
It is important to clarify the value of the switching exponent $b$ 
used to analyze the experiments 
because the value of $b$ strongly affects the evaluation of the thermal stability \cite{taniguchi11a}. 

% ===================================================================================================================================================================================== %

In this letter, 
we studied the reason why 
the linear scaling ($b=1$) seemed to work well 
to analyze the spin torque switching experiments. 
First, we showed that 
when the current is small, 
eq. (\ref{eq:Delta_general}) can be exactly rewritten 
as the switching barrier with linear scaling 
by introducing another scaling current $\tilde{I}_{\rm c}$. 
Second, we calculated the current dependence of the switching probability, 
and investigated the temperature 
and current pulse duration time regions 
in which linear scaling is valid. 
A comparison of the calculated values with the previous experiment is also discussed. 

% ===================================================================================================================================================================================== %

\begin{figure}%[p]
\centerline{\includegraphics[width=0.5\columnwidth]{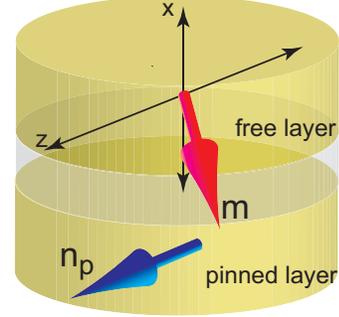}}\vspace{-3.0ex}
\caption{
        Schematic view of an in-plane magnetized system. 
        The $z$-axis is parallel to the in-plane easy axis of the free layer 
        while the $x$-axis is normal to the film plane. 
         \vspace{-3ex}}
\label{fig:fig1}
\end{figure}

% ===================================================================================================================================================================================== %

The system we consider is shown schematically in Fig. \ref{fig:fig1}, 
in which the unit vectors pointing in the magnetization directions 
of the free and pinned layers are denoted as $\mathbf{m}$ and $\mathbf{n}_{\rm p}=\mathbf{e}_{z}$, 
respectively. 
The magnetic energy density of the free layer, 
\begin{equation}
  E
  =
  -\frac{MH_{\rm K}}{2}
  m_{z}^{2}
  +
  2\pi M^{2}
  m_{x}^{2}, 
  \label{eq:energy}
\end{equation}
consists of the uniaxial anisotropy energies along the in-plane easy axis ($z$-axis) 
and the hard axis ($x$-axis) normal to the film plane. 
The energy density has two minima at $\mathbf{m}=\pm\mathbf{e}_{z}$, 
two maxima at $\mathbf{m}=\pm\mathbf{e}_{x}$, 
and two saddle points at $\mathbf{m}=\pm\mathbf{e}_{y}$. 
In the experiments, 
the external field is applied to the easy axis direction 
to quickly observe the switching \cite{yakata09}. 
However, for simplicity, 
the discussions below consider the zero applied field limit only. 

% ===================================================================================================================================================================================== %

We assume that the magnetization dynamics is described by 
the Landau-Lifshitz-Gilbert (LLG) equation 
\begin{equation}
\begin{split}
  \frac{d\mathbf{m}}{dt}
  =&
  -\gamma
  \mathbf{m}
  \times
  \mathbf{H}
  -
  \gamma
  H_{\rm s}
  \mathbf{m}
  \times
  \left(
    \mathbf{n}_{\rm p}
    \times
    \mathbf{m}
  \right)
\\
  &-
  \gamma
  \mathbf{m}
  \times
  \mathbf{h}
  +
  \alpha
  \mathbf{m}
  \times
  \frac{d\mathbf{m}}{dt},
  \label{eq:LLG}
\end{split}
\end{equation}
where $\mathbf{H}=-\partial E/\partial (M\mathbf{m})$ 
is the magnetic field. 
The strength of the spin torque $H_{\rm s}=\hbar \eta I/(2eMV)$ 
consists of the current $I$ and the spin polarization $\eta$. 
The positive current corresponds to the electron flow from the free layer to the pinned layer. 
The Gilbert damping constant and the gyromagnetic ratio are 
denoted as $\alpha$ and $\gamma$, respectively. 
The components of the random field, $\mathbf{h}$, satisfy 
the fluctuation-dissipation theorem, 
$\langle h_{i}(t)h_{j}(t^{\prime}) \rangle = [2\alpha k_{\rm B}T/(\gamma MV)]\delta_{ij}\delta(t-t^{\prime})$. 
Below, the initial state is assumed to be $\mathbf{m}=\mathbf{e}_{z}$. 

% ===================================================================================================================================================================================== %

The theoretical formula of the switching barrier is obtained 
by solving the Fokker-Planck equation in the energy space \cite{apalkov05,pinna13,taniguchi13,taniguchi13a}, 
which is derived from the statistical average of the LLG equation. 
The explicit form of the switching barrier is 
\begin{equation}
  \Delta
  =
  \frac{V}{k_{\rm B}T}
  \int_{E^{*}}^{E_{\rm s}}
  d E 
  \left(
    1
    -
    \frac{\mathscr{M}_{\rm s}}{\alpha \mathscr{M}_{\alpha}}
  \right),
  \label{eq:Delta_def}
\end{equation}
where $\mathscr{M}_{\rm s}=\gamma^{2}H_{\rm s}\oint dt [\mathbf{n}_{\rm p}\cdot\mathbf{H}-(\mathbf{m}\cdot\mathbf{n}_{\rm p})(\mathbf{m}\cdot\mathbf{H})]$ 
and $\mathscr{M}_{\alpha}=\gamma^{2}\oint dt [\mathbf{H}^{2}-(\mathbf{m}\cdot\mathbf{H})^{2}]$ are 
the functions of the energy density $E$ of the free layer, 
and are proportional to the work done by spin torque 
and the energy dissipation due to the damping 
during a precession on the constant energy line, respectively. 
%whose explicit forms are given in Refs. \cite{taniguchi13,taniguchi13a}. 
%The function $\mathscr{M}_{\rm s}$ is proportional to the current $I$. 
%The Gilbert damping constant is denoted as $\alpha$. 
The upper boundary of the integral, $E_{\rm s}$, corresponds to 
the saddle point ($\mathbf{m}=\pm\mathbf{e}_{y}$). 
On the other hand, 
the lower boundary of the integral, $E^{*}$, 
corresponds to the energy density at which 
the spin torque balances the damping. 
To discuss this point, 
the following two characteristic currents must be introduced \cite{taniguchi13,taniguchi13a}: 
\begin{equation}
  I_{\rm c}
  =
  \frac{2 \alpha e MV}{\hbar \eta}
  \left(
    H_{\rm K}
    +
    \frac{4\pi M}{2}
  \right),
  \label{eq:Ic}
\end{equation}
\begin{equation}
  I_{\rm c}^{*}
  =
  \frac{4\alpha eMV}{\pi\hbar \eta}
  \sqrt{
    4\pi M 
    \left(
      H_{\rm K}
      +
      4\pi M
    \right)
  }.
  \label{eq:Ic*}
\end{equation}
%where $\eta$ is the spin polarization of the current. 
In a thin-film geometry ($H_{\rm K} \ll 4\pi M$), 
$I_{\rm c}^{*} \simeq 1.27 I_{\rm c}$. 
The physical meanings of $I_{\rm c}$ and $I_{\rm c}^{*}$ are that 
for $I>I_{\rm c}$ the initial equilibrium state ($\mathbf{m}=\mathbf{e}_{z}$) becomes unstable 
while for $I>I_{\rm c}^{*}$ the magnetization can switch its direction 
without the thermal fluctuation \cite{taniguchi13}. 
In the following, 
we call the current regions $I < I_{\rm c}$ and $I_{\rm c} \le I < I_{\rm c}^{*}$ 
the low- and high-current regions, respectively. 
In the low-current region $I < I_{\rm c}$, 
the damping overcomes the spin torque at the equilibrium. 
In this case, $E^{*}$ is the minimum of the energy density, 
$-MH_{\rm K}/2$, 
corresponding to $\mathbf{m}=\pm\mathbf{e}_{z}$. 
Then, eq. (\ref{eq:Delta_def}) can be rewritten as 
\begin{equation}
  \Delta (I<I_{\rm c})
  = 
  \Delta_{0}
  \left(
    1
    -
    \frac{I}{\tilde{I}_{\rm c}}
  \right).
  \label{eq:Delta_low}
\end{equation}
Here, the scaling current $\tilde{I}_{\rm c}$ is defined as 
\begin{equation}
  \frac{I}{\tilde{I}_{\rm c}}
  =
  \frac{1}{MH_{\rm K}/2}
  \int_{-MH_{\rm K}/2}^{0} 
  d E 
  \frac{\mathscr{M}_{\rm s}}{\alpha \mathscr{M}_{\alpha}}.
  \label{eq:Ic_tilde_def}
\end{equation}
Using the explicit forms of $\mathscr{M}_{\rm s}$ and $\alpha \mathscr{M}_{\alpha}$ 
given in refs. \cite{taniguchi13,taniguchi13a,comment1}, 
the explicit form of $\tilde{I}_{\rm c}$ is given by 
\begin{equation}
  \tilde{I}_{\rm c}
  =
  \frac{2 \alpha eMV}{\hbar \eta}
  \frac{4\pi M}{\mathcal{S}},
  \label{eq:Ic_tilde}
\end{equation}
where the dimensionless quantity $\mathcal{S}$ is given by 
\begin{equation}
  \mathcal{S}
  =
  \int_{-\frac{k}{2}}^{0} 
  d \epsilon
  \frac{\pi (k + 2 \epsilon)}{k \sqrt{(1+k) (1-2\epsilon)} \{2\epsilon \mathsf{K}[\Bbbk(\epsilon)] + k \mathsf{E}[\Bbbk(\epsilon)]\}}, 
\end{equation}
where $k=H_{\rm K}/4\pi M$ and $\Bbbk(\epsilon)=\sqrt{(k+2\epsilon)/[k(1-2\epsilon)]}$. 
The first and second kinds of complete elliptic integrals are denoted as 
$\mathsf{K}(\Bbbk)$ and $\mathsf{E}(\Bbbk)$, respectively. 
The current $\tilde{I}_{\rm c}$ satisfies $I_{\rm c} < \tilde{I}_{\rm c} < I_{\rm c}^{*}$. 
On the other hand, 
for $I_{\rm c} \le I < I_{\rm c}^{*}$, 
$E^{*}$ satisfies 
$\mathscr{M}_{\rm s}(E^{*})=\alpha \mathscr{M}_{\alpha}(E^{*})$, 
and depends on the current. 
In that case, 
eq. (\ref{eq:Delta_def}) depends on the current nonlinearly. 

% ===================================================================================================================================================================================== %

Equation (\ref{eq:Delta_low}) means that 
by replacing $I_{\rm c}^{*}$ with $\tilde{I}_{\rm c}$, 
the switching exponent in the low-current region becomes exactly unity, 
i.e., $b$ in eq. (\ref{eq:Delta_general}) is 
\begin{equation}
  b(I<I_{\rm c})
  =
  \frac{\log(1-I/\tilde{I}_{\rm c})}{\log(1-I/I_{\rm c}^{*})}, 
\end{equation}
which satisfies $b>1$ and $\lim_{I \to 0}b=I_{\rm c}^{*}/\tilde{I}_{\rm c}$. 
We emphasize that eq. (\ref{eq:Delta_low}) is valid 
only in the low-current region 
while eq. (\ref{eq:Delta_general}) is applicable to 
the entire range of the thermally activated region. 
If the switching in the experiments occurs in the low-current region, 
the linear scaling of the switching barrier works well 
to analyze the experimental results. 
On the other hand, 
in the high-current region, 
we cannot introduce another scaling current 
that makes $b$ of eq. (\ref{eq:Delta_general}) unity. 
If the switching occurs in the high-current region, 
linear scaling is not applicable. 
Another important point indicated by eq. (\ref{eq:Delta_low}) is that, 
even if the switching occurs in the low-current region, 
the scaling current estimated using 
the linear fit of the experimentally observed $\Delta(I)$ 
is $\tilde{I}_{\rm c}$, 
not $I_{\rm c}$ nor $I_{\rm c}^{*}$. 
Since $\tilde{I}_{\rm c}<I_{\rm c}^{*}$, 
the linear fit leads to an underestimation of 
the switching current at zero temperature. 

% ===================================================================================================================================================================================== %

% ===================================================================================================================================================================================== %

\begin{figure}%[p]
\centerline{\includegraphics[width=1.0\columnwidth]{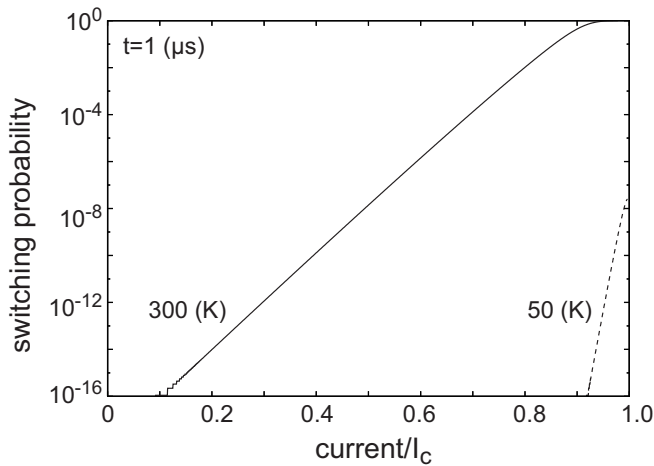}}\vspace{-3.0ex}
\caption{
        Current dependences of the switching probabilities 
        in the low-current region ($I < I_{\rm c}$) at 50 and 300 K. 
        The current pulse duration time is 1 $\mu$s. 
        The current magnitude is normalized by $I_{\rm c}$. 
         \vspace{-3ex}}
\label{fig:fig2}
\end{figure}

% ===================================================================================================================================================================================== %

% ===================================================================================================================================================================================== %

% ===================================================================================================================================================================================== %

To study whether switching occurs 
in the low-current region, 
we calculated the switching probability in the low-current region 
for several temperatures $T$ and current pulse duration times $t$. 
The switching probability is given by $P(I)=1-\exp[-ft \exp(-\Delta)]$, 
where $\Delta$ in the low-current region is given by eq. (\ref{eq:Delta_low}). 
The attempt frequency in the low-current region is \cite{taniguchi13a} 
\begin{equation}
\begin{split}
  f(I<I_{\rm c})
  =&
%  \frac{\alpha MV \mathscr{M}_{\alpha}(E_{\rm s})}{2\gamma k_{\rm B}T \tau(E^{*})}
%  \left(
%    1
%    -
%    \frac{I}{I_{\rm c}}
%  \right)
%  \left[
%    1
%    -
%    \left(
%      \frac{I}{I_{\rm c}^{*}}
%    \right)^{2}
%  \right],
  \frac{2\alpha \Delta_{0} \gamma \sqrt{4\pi M (H_{\rm K}+4\pi M)}}{\pi}
\\
  & \times
  \left(
    1
    -
    \frac{I}{I_{\rm c}}
  \right)
  \left[
    1
    -
    \left(
      \frac{I}{I_{\rm c}^{*}}
    \right)^{2}
  \right].
  \label{eq:attempt_frequency}
\end{split}
\end{equation}
%where $\mathscr{M}_{\alpha}(E_{\rm s})=4 \gamma \sqrt{H_{\rm K}4\pi M}$, 
%and $\tau(E^{*})=2\pi/[\gamma \sqrt{H_{\rm K}(H_{\rm K}+4\pi M)}]$ is 
%the precession period around the easy axis near $\mathbf{m} \simeq \mathbf{e}_{z}$. 
Figure \ref{fig:fig2} shows examples of 
the current dependences of the switching probability at 50 and 300 K, 
in which the current pulse duration time is 1 $\mu$s. 
The values of the other parameters are 
$M=1000$ emu/c.c., 
$H_{\rm K}=200$ Oe, 
$V=\pi \times 80 \times 35 \times 2.5$ nm${}^{3}$, 
$\gamma=17.64$ MHz/Oe, 
$\alpha=0.01$, 
and $\eta=0.8$, respectively, 
by which $(I_{\rm c},\tilde{I}_{\rm c},I_{\rm c}^{*})=(0.54,0.58,0.67)$ mA \cite{comment2}. 
These are typical material parameters 
of an in-plane magnetized MTJ 
consisting of CoFeB ferromagnets and MgO barrier \cite{morota08,yakata09,kubota05a,kubota05b}. 
As shown, the switching probability at 300 K reaches 
almost 100\% in the low-current region, $I<I_{\rm c}$. 
On the other hand, 
the switching probability at 50 K is much smaller than 100\%, 
which indicates that the switching at 50 K mainly occurs for $I_{\rm c} \le I < I_{\rm c}^{*}$. 

% ===================================================================================================================================================================================== %

Experimentally, 
the thermal stability $\Delta_{0}$, as well as the switching current $I_{\rm c}^{*}$, 
have been evaluated from the switching probability 
which reaches almost 100 \% \cite{morota08,yakata09,comment3}. 
When the temperature is high or the current pulse duration time is long, 
the evolution of the switching probability 
from 0 to 100\% mainly occurs in the low-current region. 
In this case, the linear scaling of the switching barrier, eq. (\ref{eq:Delta_low}), can be used 
to analyze the experimentally observed switching probability. 
The switching probability in the high-current region already reaches 100\%, 
and does not affect the evaluation of the thermal stability. 
On the other hand, when the temperature is low 
or the current pulse duration time is short, 
a large current ($<I_{\rm c}^{*}$) is required 
to saturate the switching probability 100\%. 
In this case, the evolution of the switching probability 
mainly occurs in the high-current region, 
where the linear scaling of the switching barrier is not applicable. 

% ===================================================================================================================================================================================== %

\begin{figure}%[p]
\centerline{\includegraphics[width=1.0\columnwidth]{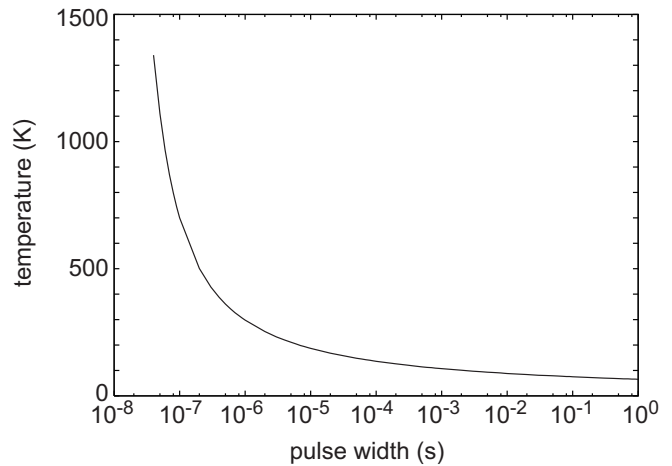}}\vspace{-3.0ex}
\caption{
         Current pulse duration time dependence of the temperature $\tilde{T}$ 
         above which the switching probability in the low-current region is larger than 99 \%. 
         The linear scaling of the switching barrier, $b=1$, 
         is applicable in the $(t,T)$ region above this line. 
         \vspace{-3ex}}
\label{fig:fig3}
\end{figure}

%To quantitatively clarify the temperature ($T$) and the current pulse duration time ($t$) regions 
%in which the linear scaling of $\Delta$ is applicable, 
%we calculated the switching probability in the low-current 
%for several $T$ and $t$ 
%by using eqs. (\ref{eq:Delta_low}) and (\ref{eq:attempt_frequency}). 
Figure \ref{fig:fig3} shows 
the relation between 
the current pulse duration time and 
the temperature $\tilde{T}$ 
above which the switching probability in the low-current region 
is larger than 99\% \cite{comment4}. 
This means that above the line in Fig. \ref{fig:fig3}, 
the switching probability reaches almost 100\% in the low-current region, 
where linear scaling of the switching barrier can be used 
to evaluate the thermal stability. 
On the other hand, below the line in Fig. \ref{fig:fig3}, 
the switching in the high-current region is not negligible, 
and linear scaling is not applicable. 
It should be noted that 
the range of the current pulse duration time in Fig. \ref{fig:fig3} is 
from 40 ns to 1 s. 
For $t<40$ ns, 
because the temperature $\tilde{T}$ is very high, 
the switching probability becomes larger than 1\% at zero current, 
%the switching probability at zero current is larger than 1 \% at the temperature $\tilde{T}$, 
which means that the initial state deviates from the easy axis significantly 
due to the large thermal fluctuation. 
%due to the high temperature. 
Therefore, we neglect the region $t<40$ ns. 
For such very short $t$, 
linear scaling is no longer applicable. 

% ===================================================================================================================================================================================== %

% ===================================================================================================================================================================================== %

% ===================================================================================================================================================================================== %

Now let us discuss the validity of 
the linear scaling of the switching barrier 
used in the analyses of the experiments. 
For example, in ref. \cite{morota08}, 
the current dependences of the switching barrier, $\Delta(I)$, 
for several current pulse duration times, $5\ \mu {\rm s} \le t \le 1\ {\rm ms}$, were measured 
at room temperature. 
The material parameters are similar to 
those used in Figs. \ref{fig:fig2} and \ref{fig:fig3}. 
According to Fig. \ref{fig:fig3}, 
$\tilde{T}$ in this range of $t$ is much lower than room temperature, 
which means that the switching in ref. \cite{morota08} occurs 
in the low-current region. 
Therefore, the linear scaling of the switching barrier is applicable 
to fit the experimental results of ref. \cite{morota08}. 
However, experimentally, 
a short current pulse 
less than 1 $\mu$s is also practicable \cite{heindl11}. 
For such short $t$, 
$\tilde{T}$ is much higher than room temperature, 
as shown in Fig. \ref{fig:fig3}, 
and therefore, 
linear scaling is not applicable. 
In this region, 
the numerical calculation of the switching rate 
from the Fokker-Planck equation is necessary 
for an accurate evaluation of the thermal stability \cite{taniguchi13a}. 

% ===================================================================================================================================================================================== %

% ===================================================================================================================================================================================== %

In summary, 
we showed a theoretical formula for the switching barrier 
of an in-plane magnetized ferromagnet, 
and pointed out that 
the switching barrier in the low-current region showed 
linear dependence on the current 
with a new scaling current $\tilde{I}_{\rm c}$. 
The temperature ($T$) and current pulse duration time ($t$) regions 
in which linear scaling of the switching barrier is applicable were obtained. 
We also pointed out that 
previous experimental analyses underestimated 
the spin torque switching current at zero temperature.

% ===================================================================================================================================================================================== %

The authors would like to acknowledge 
T. Yorozu, H. Kubota, H. Maehara, A. Emura, K. Yakushiji, H. Arai, A. Fukushima, S. Yuasa, and K. Ando 
for their valuable discussions. 
This work was supported by JSPS KAKENHI Grant-in-Aid for Young Scientists (B) 25790044. 

% ===================================================================================================================================================================================== %

%======================================================================

%====================================================================================================================================================== %


\begin{thebibliography}{99}
\bibitem{comment1} $\mathscr{M}_{\rm s}/(\alpha \mathscr{M}_{\alpha})$ in eq. (\ref{eq:Ic_tilde_def}) is $H_{\rm s}F_{\rm s}/(\alpha M F_{\alpha})$ in ref. \cite{taniguchi13} .
\bibitem{comment2} Because it is very difficult to obtain the analytical expression of $\tilde{I}_{\rm c}$,
                   the relation between $\tilde{I}_{\rm c}$ and $I_{\rm c}$, or $I_{\rm c}^{*}$, cannot be obtined analytically,
                   contrary to $I_{\rm c}^{*}/I_{\rm c} \simeq 1.27$ for $H_{\rm K} \ll 4\pi M$. 
                   Instead, we numerically evaluated the ratios, $\tilde{I}_{\rm c}/I_{\rm c}$ ($\tilde{I}_{\rm c}/I_{c}^{*}$), 
                   which varies from 1.08 to 1.05 and (from 0.85 to 0.88) for $100 \le H_{\rm K} \le 500$ Oe.
\bibitem{comment4} In other words, $\tilde{T}$ can be defined as 
                   $\lim_{I \to I_{\rm c}}P \simeq 1$. 
                   Since, strictly speaking, the theoretical switching probability $P=1-e^{-\nu t}$ is always less than 100\%, 
                   we define the switching as the probability larger than 99\%. 
                   It should also be noted that eq. (\ref{eq:attempt_frequency}) is zero at $I=I_{\rm c}$, 
                   due to limitation of the validity of the approximated solution for $f$ \cite{taniguchi13a}. 
                   Therefore, we considered the current region satisfying $dP/dI>0$ for $I<I_{\rm c}$.
\end{thebibliography}
\end{document}